\documentclass[prc,aps,onecolumn,showpacs,preprint]{revtex4}
\usepackage{graphicx}

\begin{document}

\title{Single-channel fits and K-matrix constraints
}
\author{
Ron Workman}
\affiliation{
Center for Nuclear Studies,
Department of Physics\\
The George Washington University,
Washington, D.C. 20052}

\date{\today}
 
\begin{abstract}
 
A K-matrix formalism is used to relate single-channel and
multi-channel fits. We show how the single-channel formalism
changes as new hadronic channels become accessible. These 
relations are compared to those commonly used to fit
pseudoscalar meson photoproduction data. 

\end{abstract}


\maketitle

\section{ Introduction and Motivation}

The properties of baryon resonances have largely been obtained from 
fits to $\pi N$ elastic scattering and photoproduction data. These fits have
covered the range from simple isobar models to more complex analyses involving
dispersion-relation constraints. Results have varied widely, particularly
for resonances immersed in large backgrounds and coupled only weakly to
the initial or final states. 

With the availability of high quality data involving final states
such as $\eta N$, $K \Lambda$, and $\pi \pi N$, there have been 
attempts to fit all connected reactions simultaneously, using unitarity as a 
constraint. Other groups 
have concentrated on single-channel fits to data from 
reactions such as the electro- and photoproduction of etas and kaons. The
resulting collections of masses, widths and branching fractions differ
significantly in many cases, which has led to debates over the relative importance of
particular constraints and approximations.

In the following, we examine the constraints imposed by unitarity on single-channel
fits to data using a K-matrix formalism.
Pion photoproduction will be illustrated in detail, though our
results are easily generalized to other reactions. In pion photoproduction, the
constraint imposed by Watson's theorem~\cite{watson} 
has been widely used to study the
$\Delta (1232)$ resonance region. For energies above the two-pion production
threshold, however, this constraint no longer applies. 
In an earlier work~\cite{vpi90}, we outlined a way to extend Watson's theorem
beyond the two-pion production threshold, assuming the dominance of a single
additional $\pi \Delta$ channel. This has been used in a number of subsequent
fits to pion photoproduction data. In the next section, we recall how the
form was derived and show how it reduces to Watson's theorem for an elastic
$\pi N$ scattering amplitude. We then extend this method to account for an
arbitrary number of hadronic channels. 

It is hoped that this collection of results will allow a more direct
comparison between
single-channel fits to data and multi-channel analyses based on the K-matrix
formalism. 

\section{ Results with one and two hadronic channels}

The simplest application of a K-matrix constraint to 
the two-channel process is Watson's theorem, which involves
just the channels $\pi N$ and $\gamma N$, showing that below
the two-pion production threshold, pion photoproduction amplitudes
carry the phase of the associated $\pi N$ amplitude. In Ref.~\cite{vpi90}
this argument was extended to include two hadronic channels. The second
hadronic channel was labeled $\pi \Delta$ but was intended to account for
all inelasticity in the $\pi N$ interaction. The 
$2\times 2$ hadronic K- and
T-matrices, in this case, were written as
\begin{equation}
K_H =
\left( \begin{array}{cc} K_{\pi \pi} & K_{\pi \Delta} \\
                         K_{\pi \Delta} & K_{\Delta \Delta} \end{array} \right)
\end{equation}
and
\begin{equation}
T_H =
\left( \begin{array}{cc} T_{\pi \pi} & T_{\pi \Delta} \\
                         T_{\pi \Delta} & T_{\Delta \Delta} \end{array} \right)
   \; = \; K_H ( 1 - i K_H )^{-1} ,
\end{equation}
with the abbreviations $K(\pi N\to \pi N) \equiv K_{\pi \pi}$, 
$K(\pi N\to \pi \Delta ) \equiv K_{\pi \Delta}$, and 
$K(\pi \Delta \to \pi \Delta ) \equiv K_{\Delta \Delta}$ . 

Inverting the $2\times 2$ hadronic matrix $(1-iK_H )$ and multiplying by $K_H$, we 
obtain expressions for the T-matrix elements. The T-matrix element $T_{\pi \pi}$
can be expressed, in terms of a function $\bar K$, as
\begin{equation}
T_{\pi \pi} \; = \; {{\bar K } \over {1 - i \bar K}}
\end{equation}
where
\begin{equation}
\bar K = K_{\pi \pi} + {{i K^2_{\pi \Delta} }\over {1 - i K_{\Delta \Delta} }}
\end{equation}
and
\begin{equation}
T_{\pi \Delta} = {{K_{\pi \Delta} }\over {1 - i K_{\Delta \Delta} }} 
\left( 1 + i T_{\pi \pi} \right) .
\end{equation}
The last expression obtains a factor of $(1+i T_{\pi \pi})$ through the use of
Eq.(3).

Expanding to a $3\times 3$ matrix, 
including the photon interaction channels, we have
\begin{equation}
K =
\left( \begin{array}{ccc} K_{\gamma \gamma} & K_{\gamma \pi} & K_{\gamma \Delta} \\
                         K_{\gamma \pi} & K_{\pi \pi} & K_{\pi \Delta} \\
                         K_{\gamma \Delta} & K_{\pi \Delta} & K_{\Delta \Delta}
 \end{array} \right) .
\end{equation}
Rewriting the relation $T = K ( 1 - i K )^{-1}$ for this enlarged K-matrix 
in the form
$T (1 - i K ) = K$ yields a set of relations between the T- and K-matrix
elements, a particularly useful one being,
\begin{equation}
T_{\gamma \pi} (1 - i K_{\gamma \gamma }) = (1+i T_{\pi \pi} ) K_{\gamma \pi}
               + i K_{\gamma \Delta } T_{\pi \Delta }.
\end{equation}
Using the above relation for $T_{\pi \Delta}$ and keeping terms of first order
in the electromagnetic coupling (dropping $K_{\gamma \gamma}$)
we have
\begin{equation}
T_{\gamma \pi} = (1 + i T_{\pi \pi } ) 
\left( K_{\gamma \pi} + 
i{K_{\gamma \Delta}\over {K_{\pi \Delta } }} 
{K^2_{\pi \Delta } \over {1 - i K_{\Delta \Delta } }} \right).
\end{equation}
Writing this in terms of $\bar K$ and using 
$\bar K (1 + i T_{\pi \pi}) = T_{\pi \pi}$, the final result is
\begin{equation}
T_{\gamma \pi} = (1 + i T_{\pi \pi } )
\left( K_{\gamma \pi} - {{K_{\gamma \Delta} K_{\pi \pi}}\over {K_{\pi \Delta}}} \right)
+ {K_{\gamma \Delta} \over K_{\pi \Delta}} T_{\pi \pi}.
\end{equation}

Had we proceeded with only $\gamma N$ and $\pi N$ channels, we would have
arrived at the simpler expression
\begin{equation}
T_{\gamma \pi} = (1 + i T_{\pi \pi } ) K_{\gamma \pi}
\end{equation}
which, since $T_{\pi \pi}$ is now elastic, leads directly to Watson's theorem. 
For real K-matrix elements, Eq.(9) has a phase behavior related to
$T_{\pi \pi}$ and provides a smooth connection of Watson's theorem to energies
above the two-pion production threshold. 
\vskip .5cm

\section{Extension to three or more hadronic channels}

The extension to a third hadronic channel, such as $\eta N$, can be
handled using the method described above. Writing 
$T_H (1 - i K_H ) = K_H$ for a 3-channel K-matrix, involving the
elements $\pi N$, $\pi \Delta$ and $\eta N$, again yields useful 
relations. In order to arrive at an expression similar to Eq.(9),
we retain combinations containing $T_{\pi \pi}$
\begin{equation}
T_{\pi \Delta} = {{K_{\pi \Delta} }\over {1 - i K_{\Delta \Delta} }}
\left( 1 + i T_{\pi \pi} \right) + 
{{i K_{\Delta \eta} }\over {1 - i K_{\Delta \Delta} }} T_{\pi \eta}
\end{equation}
and
\begin{equation}
T_{\pi \eta} = {1\over D} 
\left( K_{\pi \eta} + 
{{i K_{\pi \Delta} K_{\Delta \eta}}\over {1-i K_{\Delta \Delta}}} \right)
(1 + i T_{\pi \pi} )
\end{equation}
with
\begin{equation}
D= 1 - iK_{\eta \eta} + K_{\Delta \eta}^2 / (1 - i K_{\Delta \Delta} ).
\end{equation}
The $\pi N$ T-matrix element can again be represented in terms of a function
$\bar K$, as in Eq.(3), with
\begin{equation}
\bar K = K_{\pi \pi} + {{i K_{\pi \Delta}^2 }\over {1 -i K_{\Delta \Delta}}}
- {2\over D} 
{{K_{\pi \Delta} K_{\pi \eta} K_{\Delta \eta}}\over {1-iK_{\Delta \Delta}}}
- {i\over D} 
\left( {{K_{\pi \Delta} K_{\Delta \eta}}\over {1-iK_{\Delta \Delta}}} \right)^2
+ {i\over D} K_{\pi \eta}^2
\end{equation}

Substituting these relations into
\begin{equation}
T_{\gamma \pi} (1 - i K_{\gamma \gamma }) = (1+i T_{\pi \pi} ) K_{\gamma \pi}
  + i K_{\gamma \Delta } T_{\pi \Delta } +i K_{\gamma \eta} T_{\pi \eta}
\end{equation}
dropping the $K_{\gamma \gamma}$ term and regrouping, we have
\begin{equation}
T_{\gamma \pi} = (1 + i T_{\pi \pi} ) 
\left( K_{\gamma \pi} - {{K_{\gamma \Delta} K_{\pi \pi}}\over {K_{\pi \Delta}}} \right)
+ (1 + i T_{\pi \pi} )\left( A {K_{\gamma \Delta}\over K_{\pi \Delta}} 
+ B {K_{\gamma \eta}\over K_{\pi \eta}} \right)
\end{equation}
with
\begin{equation}
A = K_{\pi \pi} + {{i K_{\pi \Delta}^2 }\over {1 -i K_{\Delta \Delta}}}
- {1\over D}
{{K_{\pi \Delta} K_{\pi \eta} K_{\Delta \eta}}\over {1-iK_{\Delta \Delta}}}
- {i\over D}
\left( {{K_{\pi \Delta} K_{\Delta \eta}}\over {1-iK_{\Delta \Delta}}} \right)^2
\end{equation}
and
\begin{equation}
B = {i\over D}\left( K_{\pi \eta}^2 
+ {i {K_{\pi \Delta} K_{\pi \eta} K_{\Delta \eta}}\over {1-iK_{\Delta \Delta}}}
\right)
\end{equation}
Since $A+B=\bar K$, we can add and subtract $B K_{\gamma \Delta} / K_{\pi \Delta}$
and use $\bar K (1 + i T_{\pi \pi}) = T_{\pi \pi}$ to obtain
\begin{equation}
T_{\gamma \pi} = (1 + i T_{\pi \pi} )
\left( K_{\gamma \pi} - {{K_{\gamma \Delta} K_{\pi \pi}}\over {K_{\pi \Delta}}} \right)
+ {{K_{\gamma \Delta} }\over {K_{\pi \Delta} }} T_{\pi \pi} 
+ i \left( K_{\gamma \eta}  -
          {{K_{\gamma \Delta} }\over {K_{\pi \Delta}} } K_{\pi \eta}
\right) T_{\pi \eta}.
\end{equation} 
The extension to further hadronic channels is then obvious. Each new channel $mN$ adds
a term of the form
\begin{equation}
i \left( K_{\gamma m}  -
          {{K_{\gamma \Delta} }\over {K_{\pi \Delta}} } K_{\pi m} \right) T_{\pi m} .
\end{equation}

\vskip .5cm

\section{Results at the K-matrix pole}

It is instructive to examine these results at the K-matrix pole position. 
Taking the simplest representation involving a single pole
\begin{equation} 
K_{\gamma \pi} = { A\over {W - W_R}} + B 
\;\; ; \;\; K_{\pi \pi} = { C\over {W - W_R}} + D
\end{equation}
in the relation $T_{\gamma \pi} = K_{\gamma \pi} (1 + i T_{\pi \pi} )$, one obtains
\begin{equation}
T_{\gamma \pi} = \left( {A \over { W - W_R}} + B \right) 
 {{ ( W - W_R ) } \over { W - W_R - i [ C + ( W - W_R )D ] }}
\end{equation}
Of the two resulting terms, one is clearly `resonant' at $W$ = $W_R$ while the
other goes to zero at this energy (the `background' term). Both have the phase
of the (elastic) $\pi N$ T-matrix.

This nice correspondence disappears when a second hadronic channel is added.
With the addition of a $\pi \Delta$ channel, $(1 + i T_{\pi \pi} )$ no longer
goes to zero at $W$ = $W_R$, and the expression in Eq.(22) becomes divergent.
Assuming a pole exists in $K_{\gamma \Delta}$, both terms giving $T_{\gamma \pi}$
in Eq.(7) diverge, though the sum remains finite. 
In our Eq.(19), however, these divergences do not occur. The bracketed terms
remain finite at the K-matrix pole, assuming the residues are factorizable. 

In the simple case (Watson's theorem) involving only the $\gamma N$ and
$\pi N$ channels, there would have been no resonance term without a pole
in the K-matrix element $K_{\gamma \pi}$. However, in Eq.(9) and its generalization,
Eq.(19), the pole structures in all K-matrix 
elements multiplying the hadronic T-matrices
cancel by construction. The resonancelike behavior of $T_{\gamma \pi}$, in this case,
results from the structure of the hadronic T-matrices.

It is amusing to consider a case without explicit K-matrix poles which results
in resonancelike behavior. This is most easily seen from Eq.(4), which is
in fact the reduced K-matrix (element) of the full K-matrix involving 
$\pi N$ and $\pi \Delta$ channels. As is well known~\cite{levi}, the 
reduced K-matrix can (in principle) 
develop a pole, due to a zero in the denominator of
the second term in Eq.(4), without an explicit pole in $K_{\pi \pi}$.
This would typically occur just below the threshold for $\pi \Delta$ production,
where momentum factors implicit in $K_{\Delta \Delta}$ become complex.
However, expressions of the form 
\begin{equation}
\left( K_{\gamma \pi} - {{ K_{\gamma \Delta} } \over {K_{\pi \Delta}} }
K_{\pi \pi} \right)
\end{equation}
remain finite through the cancellation of poles at the resonance position.
The behavior of such terms would be completely different if some of the 
elements were pole free. However, if none of the K-matrix elements contained
poles, the result could be similar to one due to pole cancellation. 

\section{Single-channel versus Multi-channel fits}

We have shown how the functional forms used in fitting single-channel 
photoproduction data are related to the K-matrix elements of a multi-channel
analysis. In the $\Delta (1232)$ resonance region, Watson's theorem provides
a strong constraint on the fit. If only two hadronic channels dominate, a
form involving only the $\pi N$ T-matrix is also possible. Beyond this point,
terms proportional to other hadronic T-matrices appear to be necessary. 

It is useful to compare two single-channel forms that have been used extensively
in fitting pion photoproduction data. In the original SAID~\cite{vpi90} analysis,
data were fitted using the relation
\begin{equation}
T_{\gamma \pi} = ( {\rm Born + A} )( 1 + i T_{\pi \pi} ) + B T_{\pi \pi}
\end{equation}
wherein the `Born' term included vector meson exchanges and the terms $A$ and $B$ were
phenomenological. This corresponds to Eq.(9), with $A$ and $B$ representing
the ratios of K-matrix elements.
In the original MAID fits~\cite{maid} a simpler parametrization
\begin{equation}
T_{\gamma \pi} = ( {\rm Born} )( 1 + i T_{\pi \pi} ) + e^{i\phi} T_{\rm BW} 
\end{equation}
was used. This form explicitly separates resonance and background pieces,
the resonance part taken to be a Breit-Wigner function.
The parameter $\phi$ is included to ensure that the overall phases
of $T_{\gamma \pi}$ and $T_{\pi \pi}$ are equal at energies where Watson's theorem
is valid.

Differences between Eqs.(24) and (25) are minimized near resonances that have
small backgrounds, a clear Breit-Wigner behavior, and a large coupling to 
the $\pi N$ channel. For a nearly elastic resonance, the term $(1 + i T_{\pi \pi} )$
becomes small and masks the effect of the additional term $A$ in Eq.(24).
It should be noted, however, that many established resonances have 
$\pi N$ branching fractions of only 10-30\% . In an unbiased fit, the term $A$
can become quite large, resulting in a `background' very different from the
first term in Eq.(25). An example of this effect will be given below.

In order to make contact with multi-channel fits, comparisons should be made in
cases where the dominance of two hadronic channels is a good approximation. Here
the quantity $A$, found from fits to single-channel data, could be directly 
compared to calculated ratios of K-matrix elements. One problem with this approach
is the reliable determination of $K_{\gamma \pi}$. Missing pieces in $K_{\gamma \pi}$
could be compensated for in the phenomenological term. A consistent comparison
would require the same form of $K_{\gamma \pi}$ to be used in both the single-
and multi-channel analyses. 

Finally, we mention a reaction, other than pion
photoproduction, where the dominance of two hadronic channels is a viable
approximation. Rewriting Eq.(9) for eta photoproduction in the N(1535)
channel, we have
\begin{equation}
T_{\gamma \eta} = (1 + i T_{\eta \eta } )
\left( K_{\gamma \eta} - {{K_{\gamma \pi} K_{\eta \eta}}\over {K_{\pi \eta}}} \right)
+ {K_{\gamma \pi} \over K_{\pi \eta}} T_{\eta \eta}
\end{equation}
assuming the dominance of $\pi N$ and $\eta N$. Early single-channel fits assumed
eta photoproduction was resonance dominated with negligible background. 
The argument for this was based on the 
size of Born contributions involving a very small $\eta NN$ coupling constant.
However, while small Born terms may justify the neglect of $K_{\gamma \eta}$, 
this argument does not exclude the second term multiplying
$(1 + i T_{\eta \eta} )$. This gives a qualitative way to understand the
different results of resonance-only single-channel fits~\cite{krusche} and 
multichannel analyses~\cite{penner}. 

\section{A numerical comparison}

In Fig.~1, we compare a Born-term approximation to the sum of
Born term and phenomenological contributions multiplying $(1 + i T_{\pi \pi})$
in Eq.~(24). The phenomenological term is parameterized as a polynomial in
energy, constrained to have the proper threshold behavior. Figure 1(a) shows
the multipole $M_{1-}^{1/2}$ connected to the Roper resonance, which couples
largely to $\pi N$ and $\pi \Delta$. Figure 1(b) gives the magnetic multipole
connected to the $D_{13}(1520)$ $\pi N$ resonance. This state couples
largely to $\pi N$ and $\pi \Delta$, but also has a substantial $\rho N$ 
coupling. In an unbiased fit to data, the factors 
multiplying $(1 + i T_{\pi \pi})$
rapidly depart from a simple Born term approximation. 

The use of Born terms with point-like couplings is known to be problematic both
at threshold (where chiral perturbation theory is applicable) and at higher
energies (where form factors are often applied). With this in mind, the 
departure from a Born approximation to $K_{\gamma \pi}$ should not be
surprising. Notice that the sum of Born and phenomenological pieces changes
sign near the resonance position in these two multipoles. This behavior is 
more likely due to the phenomenological term than a modification of the Born
approximation. While the pole terms multiplying hadronic T-matrices
in Eq.~(9) cancel at $W_R$, non-pole cross terms retain a dependence on
$(W - W_R)$. These terms could account for a sign change near the resonance
energy. Not all multipoles show this cross-over behavior.
The evaluation of $K_{\gamma \Delta} K_{\pi \pi} / K_{\pi \Delta}$
within a multi-channel
model could help to clarify this issue. It should be emphasized that the
phenomenological terms used to generate the curves in Fig.~1 contain no
explicit dependence on $W_R$.  
 
\begin{acknowledgments}

R.A. Arndt is thanked for numerous discussions. We also
thank C.~Bennhold for questions that motivated this study.
This work was supported in part by the U.S. Department of
Energy Grant DE-FG02-99ER41110 and funding provided by Jefferson Lab. 
\end{acknowledgments}
\eject
    
\begin{figure*}
\centering{
\includegraphics[height=0.45\textwidth, angle=90]{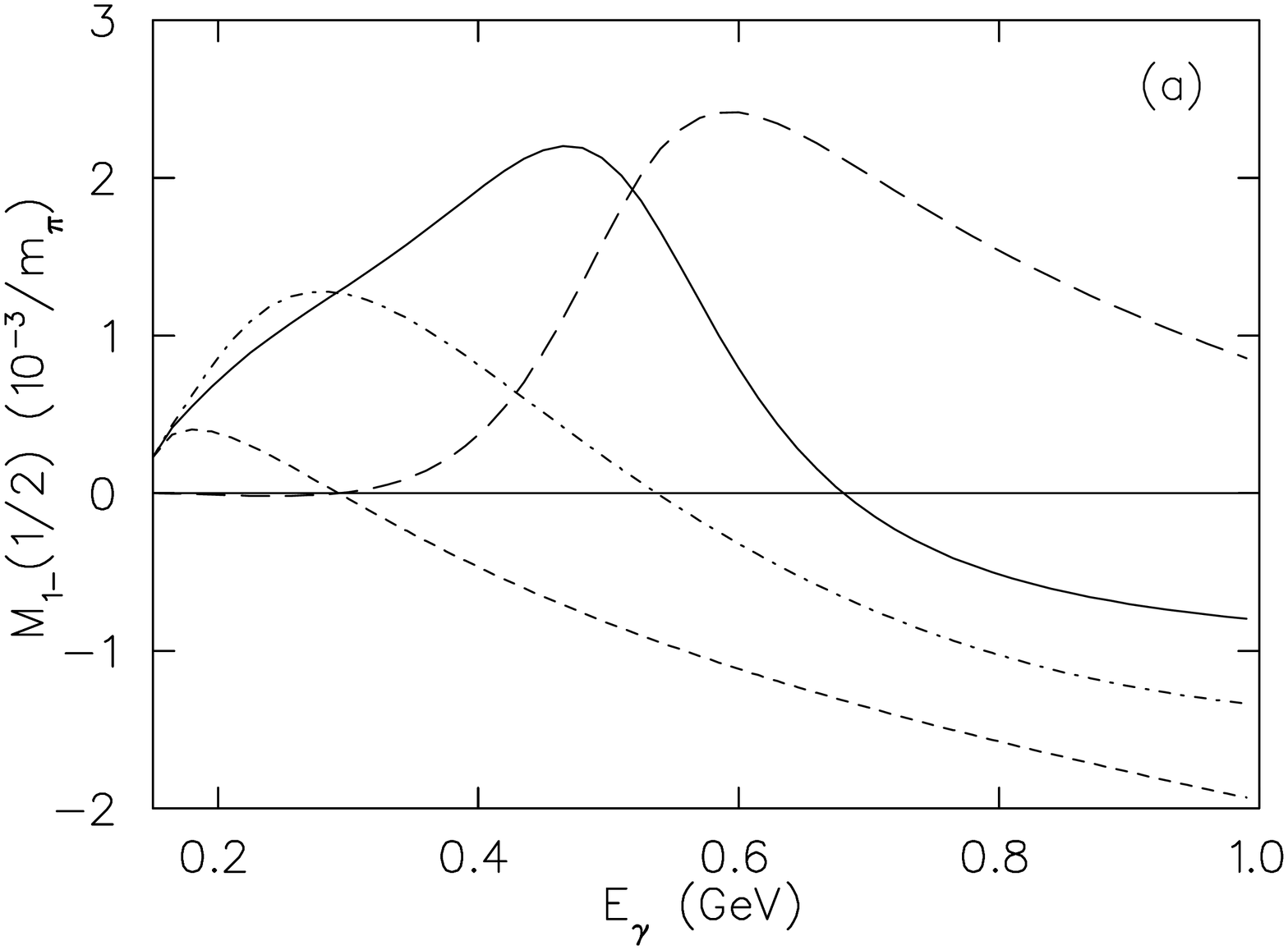}\hfill
\includegraphics[height=0.45\textwidth, angle=90]{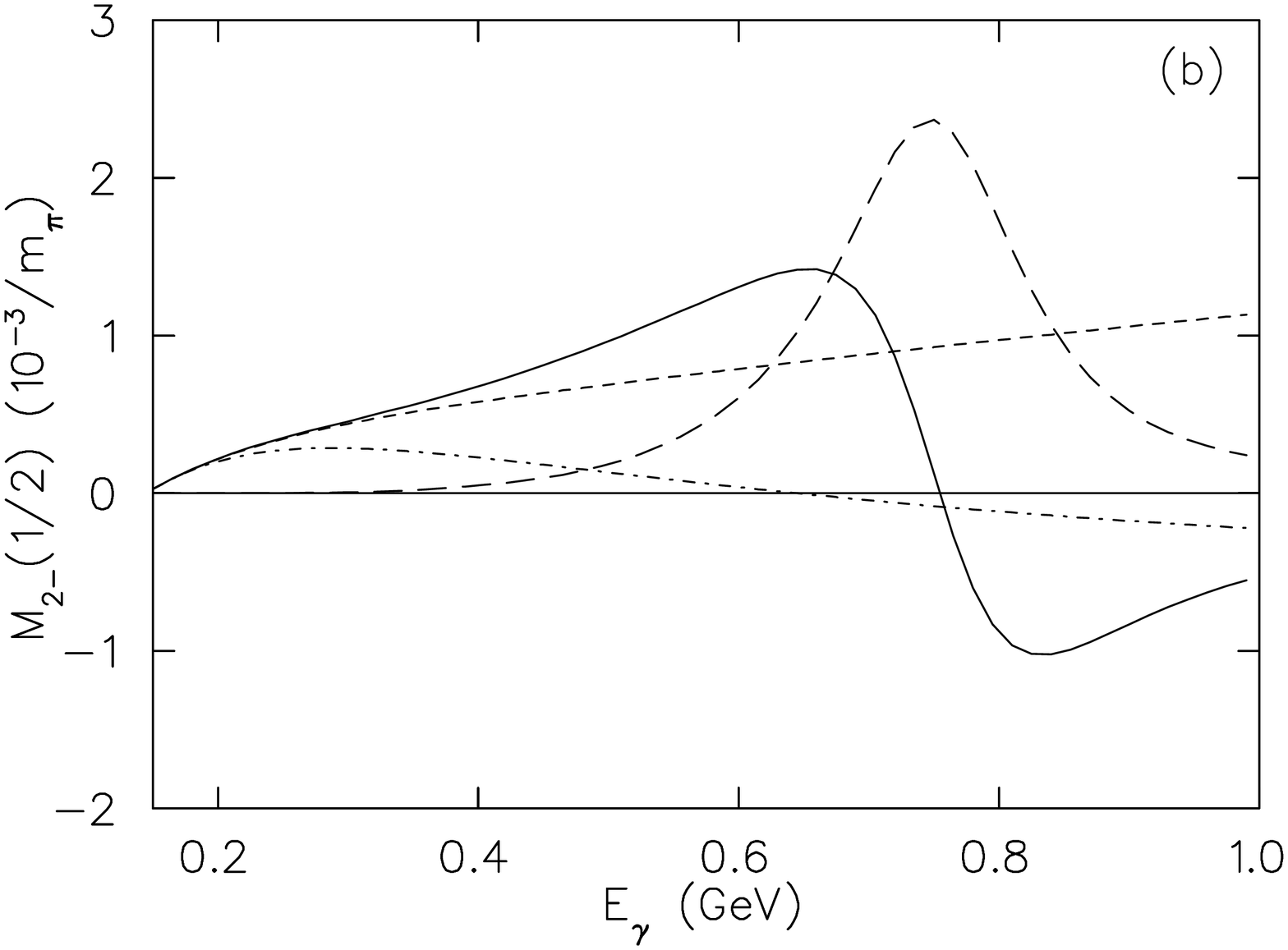}
}\caption{(a) $M_{1-}^{1/2}$ and (b) $M_{2-}^{1/2}$ multipole
          amplitudes. Solid (long dashed) curves give the
 real (imaginary) parts of the SM95~\cite{sm95} multipoles.
 The short-dashed curves give the (real) Born terms and the
 dot-dashed curves give the sum of Born + phenomenological terms.
          } \label{fig:1}
\end{figure*}

\eject

\end{document}